\documentclass[pre,twocolumn,showpacs]{revtex4}

\usepackage{amsmath,amssymb,wasysym}
\usepackage{graphicx}

\newcommand{\aver}[1]
{\left \langle #1 \right \rangle}

\newcommand{\Heff}{\mathcal{H}_{\mathrm{eff}}}

\begin{document}

\title{Nearest level spacing statistics in open chaotic systems: A generalization of the Wigner Surmise}

\author{Charles Poli}
\affiliation{Instituto de F\'isica, Universidad Aut\'onoma de Puebla, Apartado postal J-48, Puebla 72570, Mexico}

\author{Germ\'an A. Luna-Acosta}
\affiliation{Instituto de F\'isica, Universidad Aut\'onoma de Puebla, Apartado postal J-48, Puebla 72570, Mexico}

\author{Hans-J\"urgen St\"ockmann }
\affiliation{Fachbereich Physik der Philipps-Universit\"at Marburg, D-35032 Marburg, Germany}

\date{\today}

\begin{abstract}
We investigate the nearest level spacing statistics of open chaotic wave systems. To this end we derive the spacing distributions for the three Wigner ensembles in the one-channel case. The theoretical results give a clear physical meaning of the modifications on the spacing distributions produced by the coupling to the environment. Based on the analytical expressions obtained, we then propose general expressions of the spacing distributions for any number of channels, valid from weak to strong coupling. The latter expressions contain one free parameter. The surmise is successfully compared with numerical simulations of non-Hermitian random matrices and with experimental data obtained with a lossy electromagnetic chaotic cavity. 
\end{abstract}

\maketitle

It is by now well established that classical chaos mani- fests itself in generic statistical properties of the corre- sponding quantum or wave systems. In the ideal case of closed systems, i.e., those whose coupling to the environ- ment can be neglected, the spectral and spatial statistics are well described by the random matrix theory (RMT) \cite{Meh91}.  The statistics of chaotic wave systems coincide with the Gaussian orthogonal ensemble (GOE) if time-reversal symmetry (TRS) holds; with the Gaussian unitary en- semble (GUE) if TRS is broken; and with the Gaussian symplectic ensemble (GSE) for TRS systems with spin-1=2 interactions. Among all the statistical quantities used to analyze the spectral properties of closed systems, the nearest level spacing distribution PðsÞ is certainly the most referred one (see Ref. \cite{Sto99} for a review). The spacings distributions, surmised by Wigner using the 2-level approximation \cite{Wig51}, reads
\begin{equation}\label{psWig}
P_{\beta}^{\text{Wig}}(s) \propto s^\beta e^{-(A/2)s^2}\, ,
\end{equation}
where $\beta$ is the Dyson index labeling GOE ($\beta=1$), GUE ($\beta=2$), and GSE ($\beta=4$). 
These distributions were first used to describe spectral statistics of heavy nuclei and later successfully employed to describe spectral statistics of a wide range of closed and weakly open chaotic systems (see \cite{Guh98} and references therein). 

However, for systems of current interest such as quantum dots \cite{Bee97}, nuclear compounds reactions \cite{Mit10}, micro-lasers cavities \cite{Shi09, Bog10} or microwave billiards \cite{Sto99, Hem05b, Pol10, Die10},
 the coupling to the environment must be explicitly taken into account. The open systems are then characterized by a set of resonances embedded in the continuum given by the poles of the $S$-matrix  \cite{Oko03}. The poles are, in turn, the complex eigenvalues  of the effective Hamiltonian:
\begin{equation}\label{heff}
\Heff =H-\frac{i}{2}VV^\dagger \, ,
\end{equation}
where the Hermitian part $H$ is the Hamiltonian of the closed system giving rise to $N$ real energy levels and the anti-Hermitian part $iVV^\dagger/2$ models the coupling to the environment in terms of $M$ scattering channels. The $N \times M$ matrix $V$ contains the coupling amplitudes $V_n^c$ that connect the $n$th level to the $c$th scattering channel. The eigenvalues of the effective Hamiltonian are complex: $\mathcal{E}_n=E_n-\frac{i}{2}\Gamma_n$, where $E_n $ and $\Gamma_n$ are, respectively, the eigenenergies and  the resonance widths of the open system. Applying RMT to the effective Hamiltonian (\ref{heff}) (see \cite{Kuh05, Fyo05} for recent reviews), the Hermitian part of $\Heff$ is described by a Gaussian ensemble of the appropriate symmetry and the coupling amplitudes are considered as independent random Gaussian variables  \cite{Sav06}; real for GOE, complex for GUE, and real quaternion for GSE, with zero mean and covariance: $\aver{V_{n}^c(V_{n'}^{c'})^*}=(1/\eta)\delta_{nn'}\delta^{cc'}$. Here $1/\eta$ is the coupling strength.

Using the above formalism, much progress has been achieved in understanding  the statistical properties of the widths from weak \cite{Alt95} to strong coupling \cite{Som99, Kuh08} as well as the statistics of eigenvectors \cite{Sch00, Pol09b}. However, to the best of our knowledge, there are still not general analytical expressions for the distributions of the spacings $s_n=E_{n+1}-E_n$ for open chaotic systems. This is so, even if analytical results for interesting particular cases have been reported \cite{Sto98, Miz93, Sch11} and progresses have been done in  analyzing the spacing distribution of the resonances in the complex plane \cite{Gro88, Fyo97a}. What is missing then is the counterpart of the Wigner surmise for open chaotic wave systems. In this Letter, we achieve this goal by deriving analytically the probability distributions of the spacings for the 3 Gaussian ensembles for the 2-level model and the one-channel case. Then, we extend those results to the $N$-level model and to any number of channels considering the coupling strength as a free parameter. 
 
 To derive the spacing distributions $\mathcal{P}_{M=1}^{\beta}(s)$ we start with the joint energy distribution $P_{M=1}^{\beta}(\{E_n\}, \{\Gamma_n\})$, first obtained for GOE by Sokolov and Zelevinsky \cite{Sok89} and then for the 3 ensembles by St\"ockmann and \v{S}eba \cite{Sto98}. In analogy with the derivation of the Wigner surmise, we assume that the 2-level approximation holds. Specializing Eq. (4.4) of \cite{Sto98}, the joint energy distribution reads
 \begin{widetext}
\begin{equation}\label{pEG2}
\mathcal{P}_{M=1}^{\beta}(\{E_n\}, \{\Gamma_n\})\propto  \frac{ (E_1-E_2)^2+\frac{1}{4}(\Gamma_1-\Gamma_2)^2 }{\big [(E_1-E_2)^2+\frac{1}{4}(\Gamma_1+\Gamma_2)^2\big ]^{1-\beta/2 } }(\Gamma_1 \Gamma_2)^{\beta /2-1}
 \exp \bigg[-A \Big(E_1^2+E_2^2+ \frac{\Gamma_1\Gamma_2}{2} \Big) -\frac{\eta}{2}(\Gamma_{1}+ \Gamma_2) \bigg]\, ,
\end{equation}
 \end{widetext}
where $A$ fixes the mean level of the closed system: $\Delta=\pi/\sqrt{2NA}$. $\eta\Delta \gg 1$ and $ \eta\Delta \ll 1$ respectively, correspond to the weak and strong coupling regimes.  Changing to variables  $z=E_2+E_1$ and $s=E_2-E_1$ in (\ref{pEG2}) and integrating over $z$ one gets
 \begin{widetext}
\begin{equation}\label{PsG1G2}
\mathcal{P}_{M=1}^{\beta}(s,\Gamma_1,\Gamma_2)\propto \frac{ s^2+\frac{1}{4}(\Gamma_1-\Gamma_2)^2 }{\big [s^2+\frac{1}{4}(\Gamma_1+\Gamma_2)^2\big ]^{1-\beta/2 } } (\Gamma_1 \Gamma_2)^{\beta /2-1}  \exp \bigg[ -\frac{A}{2} (s^2 +\Gamma_1\Gamma_2) -\frac{\eta}{2}(\Gamma_1+\Gamma_2)\bigg]\, .
 \end{equation}
  \end{widetext}

 To go further let us consider the 3 ensembles separately. 
 
 For the GOE case, introducing in (\ref{PsG1G2}) the new variables  $x=\Gamma_1+\Gamma_2$ and  $y=\Gamma_1-\Gamma_2$, the integration over $y$ can be done. With the proper normalization constant, the distribution yields
\begin{multline}\label{PsGOEF}
\mathcal{P}_{M=1}^{\beta=1}(s)=\frac{ A\eta}{16}     e^{-\frac{A}{2}s^2}\int\limits_0^\infty dx  \frac{1}{\sqrt{s^2+\frac{x^2}{4}}}e^{-\frac{A}{16}x^2-\frac{\eta}{2}x } \\
\times \Big[\big(8s^2+x^2 \big)I_0\Big(\frac{Ax^2}{16} \Big)+x^2I_1 \Big(\frac{Ax^2}{16} \Big) \Big] \, ,
\end{multline}
where $I_n$ is the modified Bessel function of first kind.
 
For the GUE  case, the exponential prefactor can be written in terms of the partial derivatives of $A$ and $\eta$, where $\Gamma_1\Gamma_2$ and $\Gamma_1+\Gamma_2$ are generated by $-2\frac{\partial}{\partial A}$ and $-2\frac{\partial}{\partial \eta}$, respectively. The distribution can then be written as
 \begin{multline}\label{PsG1G2U}
 \mathcal{P}_{M=1}^{\beta=2}(s,\Gamma_1,\Gamma_2) \propto e^{-\frac{A}{2}s^2}  \bigg [s^2+\frac{\partial^2}{\partial \eta^2}+2\frac{\partial}{\partial A} \bigg] \\
 \times 
 e^{ -A \frac{\Gamma_1\Gamma_2}{2} -\frac{\eta}{2}(\Gamma_1+\Gamma_2)}\, .
\end{multline}
The integrations over $\Gamma_1$ and $\Gamma_2$ are now straightforward and give rise to 
\begin{multline}\label{G1G2int}
\int\limits_0^\infty d\Gamma_1\int\limits_0^\infty d\Gamma_2 \, e^{-\frac{A}{2}  \Gamma_1 \Gamma_2 -\frac{\eta}{2}(\Gamma_1 + \Gamma_2)}\, =\frac{2e^{\frac{\eta^2}{2A}}}{A}\text{E}_1 \Big(\frac{\eta^2}{2A} \Big) \, ,
\end{multline}
where $\text{E}_1(\alpha)= \int_\alpha^\infty dx \frac{e^{-x}}{x}$ is the exponential integral. Operating (\ref{G1G2int}) with $\frac{\partial^2}{\partial \eta^2}+2\frac{\partial}{\partial A} $  and including  the normalization constant we finally arrive at
\begin{equation}\label{PsGUEF}
\mathcal{P}_{M=1}^{\beta=2}(s) = \sqrt{\frac{A}{2\pi}}\eta^2 \bigg[E(A,\eta)\, s^2+\frac{2}{\eta ^2}-\frac{E(A,\eta)}{A} \bigg ] e^{-\frac{A}{2}s^2}\, ,
\end{equation}
where $E(A,\eta)=e^{\frac{\eta^2}{2A}}\text{E}_1 \big(\frac{\eta^2}{2A} \big)$.
Applying exactly the same method to GSE, one gets
\begin{widetext}
\begin{eqnarray}\label{PsGSEF}
\mathcal{P}_{M=1}^{\beta=4}(s)= \sqrt{\frac{2\pi}{A}}\frac{A^3}{3\pi}\frac{\eta^4}{2^3}\bigg[s^4 \bigg ( -\frac{2}{A^2}+\frac{\eta^2+2A}{A^3} E(A,\eta) \bigg)
+s^2 \bigg ( -\frac{\eta^4+4\eta^2A-4A^2}{\eta^2A^4} +(\eta^2+6A)\frac{\eta^2}{2A^5}E(A,\eta) \bigg) \nonumber \\
+\frac{\eta^6+7\eta^4A-4\eta^2A^2+12A^3}{\eta^4A^5} -\frac{\eta^4+9\eta^2A+6A^2}{2A^6}E(A,\eta) \bigg ] e^{-\frac{A}{2}s^2} \, .
\end{eqnarray}
\end{widetext}

 \begin{figure*}
\begin{center}
	\includegraphics[width=1.9in]{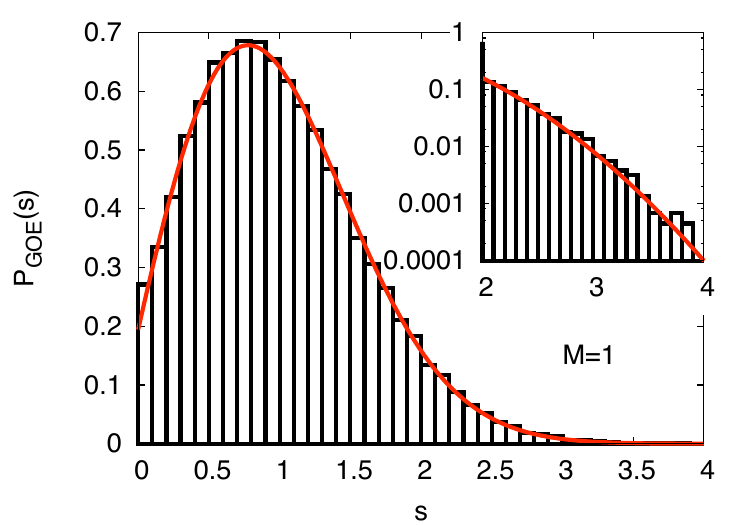}
	\includegraphics[width=1.9in]{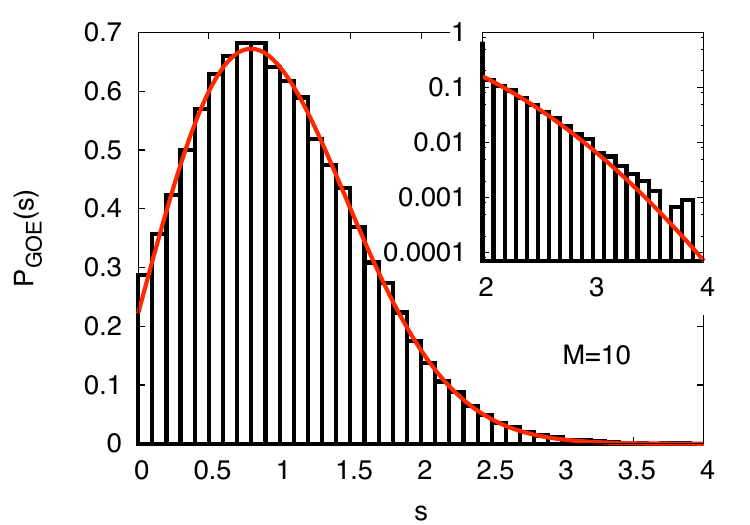}
	
	\includegraphics[width=1.9in]{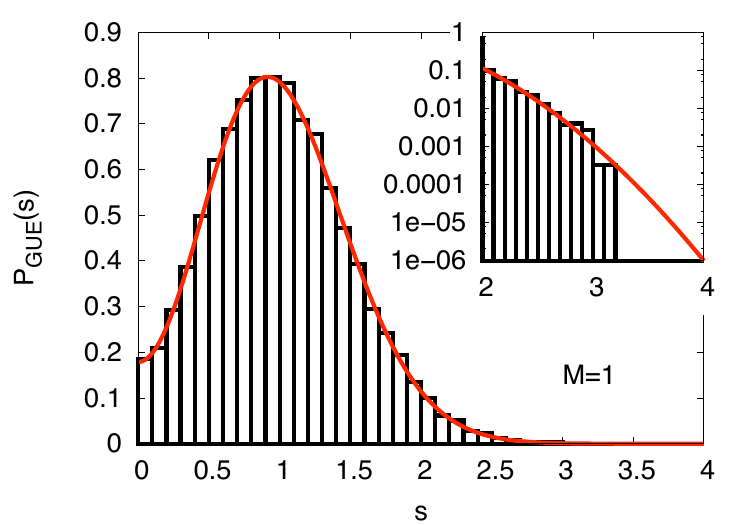}
	\includegraphics[width=1.9in]{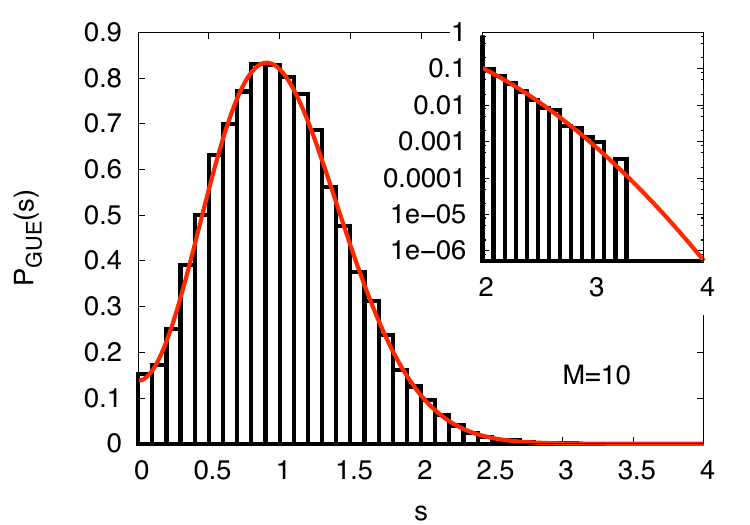}
	
	\includegraphics[width=1.9in]{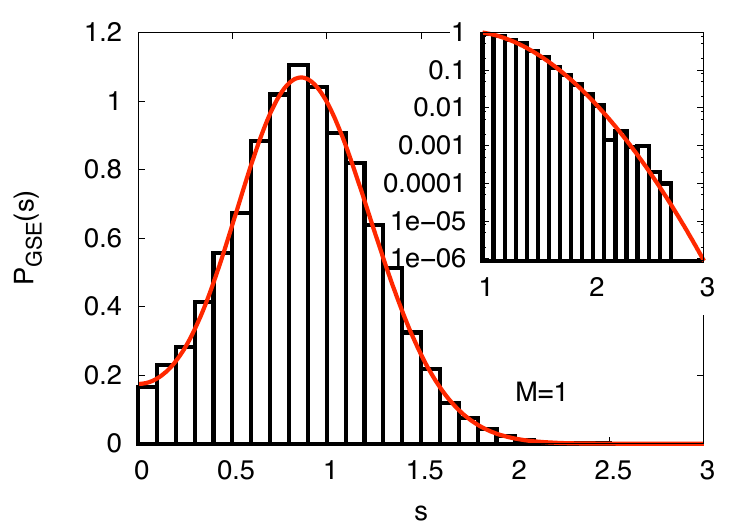}
	\includegraphics[width=1.9in]{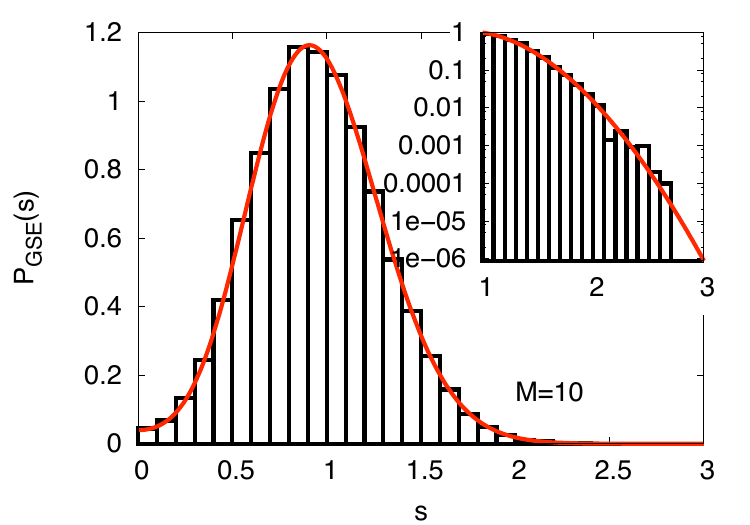}	
	\caption{Distribution of the normalized spacings for the  3 Gaussian ensembles with $\aver{\Gamma}/\Delta=1$. On the top for GOE, on the middle for GUE, and on the bottom for GSE. On the left for $M=1$ and on the right for $M=10$. For the one-channel case, the exact analytical expressions (\ref{PsGOEF}), (\ref{PsGUEF}) and (\ref{PsGSEF}) are represented; for $M=10$, considering $1/\eta$ as a free parameter. In insert the tail of the distributions are represented in semi-log scale. For all curves, the mean level spacing is normalized to one.}\label{ps}
	\end{center} 
\end{figure*}

Note that the distributions (\ref{PsGOEF}), (\ref{PsGUEF}) and (\ref{PsGSEF}) tend to the Wigner surmise (\ref{psWig}) in the limit of vanishing coupling $1/\eta \rightarrow 0$. The Gaussian tail $e^{-As^2/2}$, proper of the closed systems  (\ref{psWig}), remains and the Wigner prefactor $s^\beta$ now becomes some function $f_\beta$ of $s$: $\mathcal{P}_{M=1}^{\beta}(s)=f_\beta(s)e^{-(A/2)s^2}$. These expressions reveal also that the effects of the coupling show up only in the prefactor.  Thus, the main modifications  occur at small spacings.

In particular, there appears a finite probability for null spacings: $\mathcal{P}_{M=1}^{\beta}(s)\neq 0$. In other words, the level repulsion, a main feature of the Wigner surmise, is suppressed for open systems. This effect is associated to the phenomenon of attraction of the levels along the real axis induced by the coupling to the environment  \cite{Pol09a}. For the GUE case $f_{\beta=2}$ is a polynomial containing the Wigner term $s^2$ and a constant term. For the GSE case, in addition to the Wigner term $s^4$ and the constant term, a  quadratic term appears too. The presence of this quadratic term can be viewed as a consequence of the breaking of the TRS, since it characterizes the behavior at small spacings for closed systems with broken TRS.

It is important to note that in the limit of strong coupling $1/\eta \rightarrow \infty$ we obtain a Gaussian distribution $\mathcal{P}_{M=1}^{\beta}(s) \rightarrow\sqrt{\frac{2A}{\pi}}e^{-(A/2)s^2}$. This behavior, intrinsic of the 2-level approximation, does not correspond to the $N$-level model where the Wigner surmise is recovered \cite{Sto98}. Indeed, the 2-level model expressions are valid only up to the intermediate coupling regime. However, we will show that if $1/\eta$ is considered now as a free parameter \textit{i.e.}, an effective coupling strength called $1/\eta_{\text{eff}}$, the distributions (\ref{PsGOEF}), (\ref{PsGUEF}) and (\ref{PsGSEF}) are valid for any coupling strength and, moreover, for any number of channels.  
  To confirm the validity of our proposal and to analyze how $1/\eta_{\text{eff}}$ evolves with the coupling and the number of channels, we carried out numerical simulations of non-Hermitian random matrices defined by (\ref{heff}). The effective coupling strength $1/\eta_{\text{eff}}$ was determined by a least-square algorithm; $A$ being fixed by the usual normalization condition $\aver{s}=1$. The simulations were performed with ensembles of 150 non-Hermitian of size $1000\times 1000$ varying the mean width $\aver{\Gamma}$ in the range $[0.1,\, 30]\,  \Delta $ and for $M=1$, 3, 5 and 10.  
 
 A comparison between the expressions (\ref{PsGOEF}), (\ref{PsGUEF}) and (\ref{PsGSEF})  and the numerical results is presented in Fig. \ref{ps}. For all coupling values and number of channels analyzed the confidence level is larger than 99.5\%.  We stress that the excellent agreement with the numerical simulations confirms that the Wigner generalizations Eqs. (\ref{PsGOEF}), (\ref{PsGUEF}) and (\ref{PsGSEF}), being exact for the 2-level model and $M=1$, are excellent effective formulas for arbitrary coupling strength, number of levels and channels.

\begin{figure}[t]
\begin{center}
	\includegraphics[width=2.2in]{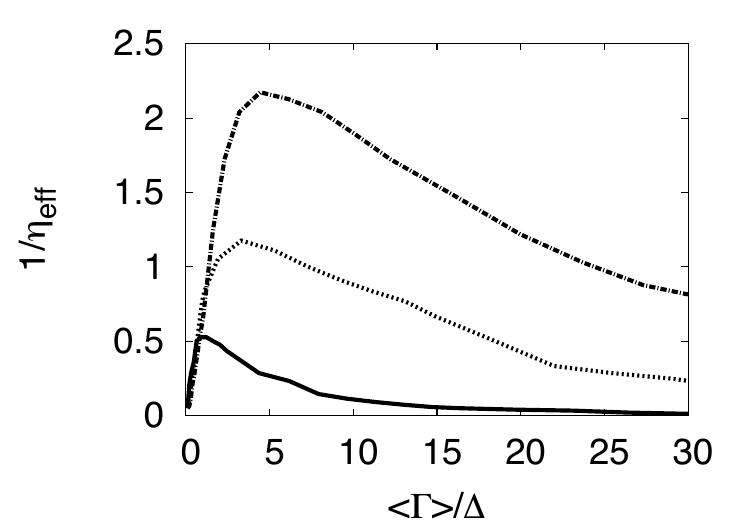}
	\caption{Evolution of the effective coupling strength $1/\eta_{\text{eff}}$ as a function of the normalized mean width $\aver{\Gamma}/\Delta$ for the GOE case. $M=1$ in solid line, $M=3$ in dots, and $M=5$ in dashed line.}\label{parameters}
	\end{center} 
\end{figure}

Using the best fit procedure  mentioned above for the effective coupling strength, we present in Fig. \ref{parameters} the evolution of  $1/\eta_{\text{eff}}$ for the GOE case as a function of the mean width. We can distinguish two regimes. For weak enough coupling ($\aver{\Gamma} \apprle 2 \Delta$), $1/\eta_{\text{eff}}$ increases quite rapidly with the mean width. For strong coupling, the reverse occurs and we go back to the Wigner surmise as predicted in \cite{Sto98} for the one-channel case. Note that the approach to the Wigner surmise also occurs for $M>1$  (and $M\ll N$), albeit extremely slow. A similar behavior was observed for the GUE and GSE cases.

\begin{figure}[h!]
\begin{center}
	\includegraphics[width=2.2in]{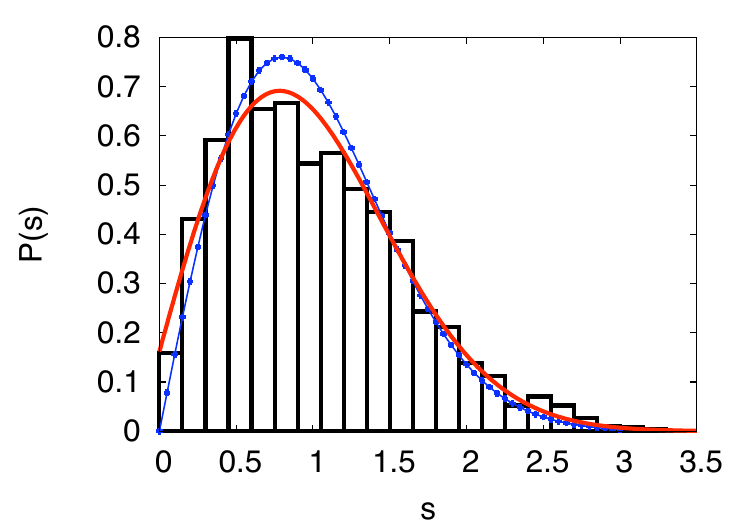}
	\includegraphics[width=2.2in]{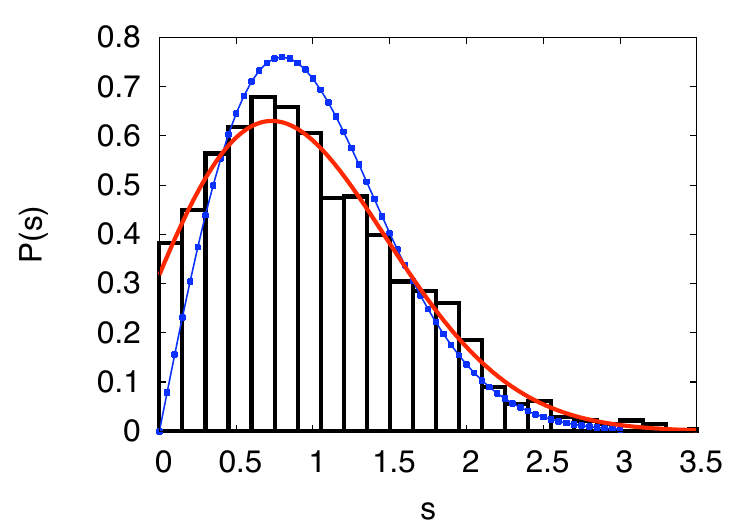}
	\caption{Distribution of the spacings for an open chaotic microwave cavity. In histogram the experimental distribution, on the top with the range 1 to 6 GHz ($M_w=10$) and on the bottom 15 to 16 GHz ($M_w=20$). Both histograms have been obtained using 2600 events. The dotted line shows the Wigner surmise, and the solid line the best fit using expression \ref{PsGOEF}.  The confidence level is lower than 50\% for the Wigner surmise (\ref{psWig}) and larger than 90\% for the expression (\ref{PsGOEF}).}\label{Ulle}
	\end{center} 
\end{figure}

Furthermore, we compare the theoretical prediction for GOE, Eq. (\ref{PsGOEF}), with recent experimental data obtained for an open chaotic 2D microwave cavity displaying TRS in the closed limit. The data were provided to us kindly by U. Kuhl \cite{Kuh08}. In that experiment, the coupling to the environment is produced by the finite conductivity of the wall (modeled by $M_w$ fictitious open channels) and one antenna attached to the cavity to perform the measurements. Fig. \ref{Ulle} shows the experimental distribution of the spacings for two different ranges of frequency \textit{i.e.}, for two different coupling strengths and number of coupling channels. The agreement is good with a confidence level larger that $90\%$ in both cases.  

In summary, we have proposed general expressions for the distributions of the nearest level spacings of open chaotic wave systems, where the internal Hamiltonian belongs to one of the three well-known classes of Gaussian Ensembles. We stress again that the expressions are exact for $N=2$ and $M=1$ and otherwise excellent approximations considering the coupling strength as a free parameter. 
The expressions indicated that coupling to the environment modified, with respect to the Wigner surmise,  the form of the cofactors from $s^\beta$ to some function of $s$. Our predictions, valid from the weak to the strong coupling regime were shown to be in excellent agreement with numerical simulations of non-Hermitian random matrices and with experimental data from a microwave cavity. In essence, the set of distributions (\ref{PsGOEF}), (\ref{PsGUEF}) and (\ref{PsGSEF}) may be viewed as the extension of the Wigner surmise to open chaotic systems. 
\begin{acknowledgments}

We thank, U. Kuhl who supplied to us the experimental data of the microwave cavity, R. Mendez-Sanchez and D. V. Savin for fruitful discussions and T. H. Seligman and the Centro International de Ciencias (Cuernavaca) where the idea to this paper was developed.
G. A. L.-A. acknowledges partial support from CONACYT Ref.  No. 51458. 
\end{acknowledgments}


\begin{thebibliography}{27}
\expandafter\ifx\csname natexlab\endcsname\relax\def\natexlab#1{#1}\fi
\expandafter\ifx\csname bibnamefont\endcsname\relax
  \def\bibnamefont#1{#1}\fi
\expandafter\ifx\csname bibfnamefont\endcsname\relax
  \def\bibfnamefont#1{#1}\fi
\expandafter\ifx\csname citenamefont\endcsname\relax
  \def\citenamefont#1{#1}\fi
\expandafter\ifx\csname url\endcsname\relax
  \def\url#1{\texttt{#1}}\fi
\expandafter\ifx\csname urlprefix\endcsname\relax\def\urlprefix{URL }\fi
\providecommand{\bibinfo}[2]{#2}
\providecommand{\eprint}[2][]{\url{#2}}

\bibitem[{\citenamefont{Mehta}(2004)}]{Meh91}
\bibinfo{author}{\bibfnamefont{M.~L.} \bibnamefont{Mehta}},
  \emph{\bibinfo{title}{Random Matrices}}
  (\bibinfo{publisher}{Elsevier, San Diego, USA}, \bibinfo{year}{2004}).

\bibitem[{\citenamefont{St\"ockmann}(1999)}]{Sto99}
\bibinfo{author}{\bibfnamefont{H.-J.} \bibnamefont{St\"ockmann}},
  \emph{\bibinfo{title}{{Quantum Chaos: an introduction}}}
  (\bibinfo{publisher}{Cambridge University Press, Cambridge, U. K.},
  \bibinfo{year}{1999}).

\bibitem[{\citenamefont{Wigner}(1951)}]{Wig51}
\bibinfo{author}{\bibfnamefont{E.~P.} \bibnamefont{Wigner}}, in
  \emph{\bibinfo{booktitle}{Proc. Cambridge Philos. Soc}}
  (\bibinfo{year}{1951}), vol.~\bibinfo{volume}{47}, p. \bibinfo{pages}{790}.

\bibitem[{\citenamefont{Guhr et~al.}(1998)\citenamefont{Guhr,
  M\"uller-Groeling, and Weidenm\"uller}}]{Guh98}
\bibinfo{author}{\bibfnamefont{T.}~\bibnamefont{Guhr}},
  \bibinfo{author}{\bibfnamefont{A.}~\bibnamefont{M\"uller-Groeling}},
  \bibnamefont{and} \bibinfo{author}{\bibfnamefont{H.~A.}
  \bibnamefont{Weidenm\"uller}}, \bibinfo{journal}{Phys. Rep.}
  \textbf{\bibinfo{volume}{299}}, \bibinfo{pages}{189} (\bibinfo{year}{1998}).

\bibitem[{\citenamefont{Beenakker}(1997)}]{Bee97}
\bibinfo{author}{\bibfnamefont{C.~W.~J.} \bibnamefont{Beenakker}},
  \bibinfo{journal}{Rev. Mod. Phys.} \textbf{\bibinfo{volume}{69}},
  \bibinfo{pages}{731} (\bibinfo{year}{1997}).

\bibitem[{\citenamefont{Mitchell et~al.}(2010)\citenamefont{Mitchell, Richter,
  and Weidenm\"uller}}]{Mit10}
\bibinfo{author}{\bibfnamefont{G.~E.} \bibnamefont{Mitchell}},
  \bibinfo{author}{\bibfnamefont{A.}~\bibnamefont{Richter}}, \bibnamefont{and}
  \bibinfo{author}{\bibfnamefont{H.~A.} \bibnamefont{Weidenm\"uller}},
  \bibinfo{journal}{Rev. Mod. Phys.} \textbf{\bibinfo{volume}{82}},
  \bibinfo{pages}{2845} (\bibinfo{year}{2010}).

\bibitem[{\citenamefont{Shinohara et~al.}(2009)\citenamefont{Shinohara,
  Hentschel, Wiersig, Sasaki, and Harayama}}]{Shi09}
\bibinfo{author}{\bibfnamefont{S.}~\bibnamefont{Shinohara}},
  \bibinfo{author}{\bibfnamefont{M.}~\bibnamefont{Hentschel}},
  \bibinfo{author}{\bibfnamefont{J.}~\bibnamefont{Wiersig}},
  \bibinfo{author}{\bibfnamefont{T.}~\bibnamefont{Sasaki}}, \bibnamefont{and}
  \bibinfo{author}{\bibfnamefont{T.}~\bibnamefont{Harayama}},
  \bibinfo{journal}{Phys. Rev. A} \textbf{\bibinfo{volume}{80}},
  \bibinfo{pages}{031801} (\bibinfo{year}{2009}).

\bibitem[{\citenamefont{Bogomolny et~al.}(2011)\citenamefont{Bogomolny,
  Djellali, Dubertrand, Gozhyk, Lebental, Schmit, Ulysse, and Zyss}}]{Bog10}
\bibinfo{author}{\bibfnamefont{E.}~\bibnamefont{Bogomolny}},
  \bibinfo{author}{\bibfnamefont{N.}~\bibnamefont{Djellali}},
  \bibinfo{author}{\bibfnamefont{R.}~\bibnamefont{Dubertrand}},
  \bibinfo{author}{\bibfnamefont{I.}~\bibnamefont{Gozhyk}},
  \bibinfo{author}{\bibfnamefont{M.}~\bibnamefont{Lebental}},
  \bibinfo{author}{\bibfnamefont{C.}~\bibnamefont{Schmit}},
  \bibinfo{author}{\bibfnamefont{C.}~\bibnamefont{Ulysse}}, \bibnamefont{and}
  \bibinfo{author}{\bibfnamefont{J.}~\bibnamefont{Zyss}},
  \bibinfo{journal}{Phys. Rev. E} \textbf{\bibinfo{volume}{83}},
  \bibinfo{pages}{036208} (\bibinfo{year}{2011}).

\bibitem[{\citenamefont{Hemmady et~al.}(2005)\citenamefont{Hemmady, Zheng,
  Antonsen, Ott, and Anlage}}]{Hem05b}
\bibinfo{author}{\bibfnamefont{S.}~\bibnamefont{Hemmady}},
  \bibinfo{author}{\bibfnamefont{X.}~\bibnamefont{Zheng}},
  \bibinfo{author}{\bibfnamefont{T.~M.} \bibnamefont{Antonsen}},
  \bibinfo{author}{\bibfnamefont{E.}~\bibnamefont{Ott}}, \bibnamefont{and}
  \bibinfo{author}{\bibfnamefont{S.~M.} \bibnamefont{Anlage}},
  \bibinfo{journal}{Phys. Rev. E} \textbf{\bibinfo{volume}{71}},
  \bibinfo{pages}{056215} (\bibinfo{year}{2005}).

\bibitem[{\citenamefont{Poli et~al.}(2010)\citenamefont{Poli, Legrand, and
  Mortessagne}}]{Pol10}
\bibinfo{author}{\bibfnamefont{C.}~\bibnamefont{Poli}},
  \bibinfo{author}{\bibfnamefont{O.}~\bibnamefont{Legrand}}, \bibnamefont{and}
  \bibinfo{author}{\bibfnamefont{F.}~\bibnamefont{Mortessagne}},
  \bibinfo{journal}{Phys. Rev. E} \textbf{\bibinfo{volume}{82}},
  \bibinfo{pages}{055201(R)} (\bibinfo{year}{2010}).

\bibitem[{\citenamefont{Dietz et~al.}(2010)\citenamefont{Dietz, Friedrich,
  Harney, Miski-Oglu, Richter, Sch\"afer, and Weidenm\"uller}}]{Die10}
\bibinfo{author}{\bibfnamefont{B.}~\bibnamefont{Dietz}},
  \bibinfo{author}{\bibfnamefont{T.}~\bibnamefont{Friedrich}},
  \bibinfo{author}{\bibfnamefont{H.~L.} \bibnamefont{Harney}},
  \bibinfo{author}{\bibfnamefont{M.}~\bibnamefont{Miski-Oglu}},
  \bibinfo{author}{\bibfnamefont{A.}~\bibnamefont{Richter}},
  \bibinfo{author}{\bibfnamefont{F.}~\bibnamefont{Sch\"afer}},
  \bibnamefont{and} \bibinfo{author}{\bibfnamefont{H.~A.}
  \bibnamefont{Weidenm\"uller}}, \bibinfo{journal}{Phys. Rev. E}
  \textbf{\bibinfo{volume}{81}}, \bibinfo{pages}{036205}
  (\bibinfo{year}{2010}).

\bibitem[{\citenamefont{Okolowicz et~al.}(2003)\citenamefont{Okolowicz,
  Ploszajczak, and Rotter}}]{Oko03}
\bibinfo{author}{\bibfnamefont{J.}~\bibnamefont{Okolowicz}},
  \bibinfo{author}{\bibfnamefont{M.}~\bibnamefont{Ploszajczak}},
  \bibnamefont{and} \bibinfo{author}{\bibfnamefont{I.}~\bibnamefont{Rotter}},
  \bibinfo{journal}{Phys. Rep.} \textbf{\bibinfo{volume}{374}},
  \bibinfo{pages}{271} (\bibinfo{year}{2003}).

\bibitem[{\citenamefont{Kuhl et~al.}(2005)\citenamefont{Kuhl, St\"ockmann, and
  Weaver}}]{Kuh05}
\bibinfo{author}{\bibfnamefont{U.}~\bibnamefont{Kuhl}},
  \bibinfo{author}{\bibfnamefont{H.-J.} \bibnamefont{St\"ockmann}},
  \bibnamefont{and} \bibinfo{author}{\bibfnamefont{R.~L.}
  \bibnamefont{Weaver}}, \bibinfo{journal}{J. Phys. A}
  \textbf{\bibinfo{volume}{38}}, \bibinfo{pages}{10433} (\bibinfo{year}{2005}).

\bibitem[{\citenamefont{Fyodorov et~al.}(2005)\citenamefont{Fyodorov, Savin,
  and Sommers}}]{Fyo05}
\bibinfo{author}{\bibfnamefont{Y.~V.} \bibnamefont{Fyodorov}},
  \bibinfo{author}{\bibfnamefont{D.~V.} \bibnamefont{Savin}}, \bibnamefont{and}
  \bibinfo{author}{\bibfnamefont{H.-J.} \bibnamefont{Sommers}},
  \bibinfo{journal}{J. Phys. A} \textbf{\bibinfo{volume}{38}},
  \bibinfo{pages}{10731} (\bibinfo{year}{2005}).

\bibitem[{\citenamefont{Savin et~al.}(2006)\citenamefont{Savin, Legrand, and
  Mortessagne}}]{Sav06}
\bibinfo{author}{\bibfnamefont{D.~V.} \bibnamefont{Savin}},
  \bibinfo{author}{\bibfnamefont{O.}~\bibnamefont{Legrand}}, \bibnamefont{and}
  \bibinfo{author}{\bibfnamefont{F.}~\bibnamefont{Mortessagne}},
  \bibinfo{journal}{Europhys. Lett.} \textbf{\bibinfo{volume}{76}},
  \bibinfo{pages}{774} (\bibinfo{year}{2006}).

\bibitem[{\citenamefont{Alt et~al.}(1995)\citenamefont{Alt, Gr\"af, Harney,
  Hofferbert, Lengeler, Richter, Schardt, and Weidenm\"uller}}]{Alt95}
\bibinfo{author}{\bibfnamefont{H.}~\bibnamefont{Alt}},
  \bibinfo{author}{\bibfnamefont{H.~D.} \bibnamefont{Gr\"af}},
  \bibinfo{author}{\bibfnamefont{H.}~\bibnamefont{Harney}},
  \bibinfo{author}{\bibfnamefont{R.}~\bibnamefont{Hofferbert}},
  \bibinfo{author}{\bibfnamefont{H.}~\bibnamefont{Lengeler}},
  \bibinfo{author}{\bibfnamefont{A.}~\bibnamefont{Richter}},
  \bibinfo{author}{\bibfnamefont{P.}~\bibnamefont{Schardt}}, \bibnamefont{and}
  \bibinfo{author}{\bibfnamefont{H.~A.} \bibnamefont{Weidenm\"uller}},
  \bibinfo{journal}{Phys. Rev. Lett.} \textbf{\bibinfo{volume}{74}},
  \bibinfo{pages}{62} (\bibinfo{year}{1995}).

\bibitem[{\citenamefont{Sommers et~al.}(1999)\citenamefont{Sommers, Fyodorov,
  and Titov}}]{Som99}
\bibinfo{author}{\bibfnamefont{H.-J.} \bibnamefont{Sommers}},
  \bibinfo{author}{\bibfnamefont{Y.~V.} \bibnamefont{Fyodorov}},
  \bibnamefont{and} \bibinfo{author}{\bibfnamefont{M.}~\bibnamefont{Titov}},
  \bibinfo{journal}{J. Phys. A} \textbf{\bibinfo{volume}{32}},
  \bibinfo{pages}{L77} (\bibinfo{year}{1999}).

\bibitem[{\citenamefont{Kuhl et~al.}(2008)\citenamefont{Kuhl, H\"ohmann, Main,
  and St\"ockmann}}]{Kuh08}
\bibinfo{author}{\bibfnamefont{U.}~\bibnamefont{Kuhl}},
  \bibinfo{author}{\bibfnamefont{R.}~\bibnamefont{H\"ohmann}},
  \bibinfo{author}{\bibfnamefont{J.}~\bibnamefont{Main}}, \bibnamefont{and}
  \bibinfo{author}{\bibfnamefont{H.-J.} \bibnamefont{St\"ockmann}},
  \bibinfo{journal}{Phys. Rev. Lett.} \textbf{\bibinfo{volume}{100}},
  \bibinfo{pages}{254101} (\bibinfo{year}{2008}).

\bibitem[{\citenamefont{Schomerus et~al.}(2000)\citenamefont{Schomerus, Frahm,
  Patra, and Beenakker}}]{Sch00}
\bibinfo{author}{\bibfnamefont{H.}~\bibnamefont{Schomerus}},
  \bibinfo{author}{\bibfnamefont{K.~M.} \bibnamefont{Frahm}},
  \bibinfo{author}{\bibfnamefont{M.}~\bibnamefont{Patra}}, \bibnamefont{and}
  \bibinfo{author}{\bibfnamefont{C.~W.~J.} \bibnamefont{Beenakker}},
  \bibinfo{journal}{Physica A} \textbf{\bibinfo{volume}{278}},
  \bibinfo{pages}{469} (\bibinfo{year}{2000}).

\bibitem[{\citenamefont{Poli et~al.}(2009{\natexlab{a}})\citenamefont{Poli,
  Savin, Legrand, and Mortessagne}}]{Pol09b}
\bibinfo{author}{\bibfnamefont{C.}~\bibnamefont{Poli}},
  \bibinfo{author}{\bibfnamefont{D.~V.} \bibnamefont{Savin}},
  \bibinfo{author}{\bibfnamefont{O.}~\bibnamefont{Legrand}}, \bibnamefont{and}
  \bibinfo{author}{\bibfnamefont{F.}~\bibnamefont{Mortessagne}},
  \bibinfo{journal}{Phys. Rev. E} \textbf{\bibinfo{volume}{80}},
  \bibinfo{pages}{046203} (\bibinfo{year}{2009}{\natexlab{a}}).

\bibitem[{\citenamefont{St\"ockmann and \v{S}eba}(1998)}]{Sto98}
\bibinfo{author}{\bibfnamefont{H.}~\bibnamefont{St\"ockmann}} \bibnamefont{and}
  \bibinfo{author}{\bibfnamefont{P.}~\bibnamefont{\v{S}eba}},
  \bibinfo{journal}{J. Phys. A} \textbf{\bibinfo{volume}{31}},
  \bibinfo{pages}{3439} (\bibinfo{year}{1998}).

\bibitem[{\citenamefont{Mizutori and Zelevinsky}(1993)}]{Miz93}
\bibinfo{author}{\bibfnamefont{S.}~\bibnamefont{Mizutori}} \bibnamefont{and}
  \bibinfo{author}{\bibfnamefont{V.~G.} \bibnamefont{Zelevinsky}},
  \bibinfo{journal}{Z. Phys. A} \textbf{\bibinfo{volume}{346}},
  \bibinfo{pages}{1} (\bibinfo{year}{1993}).

\bibitem[{\citenamefont{Shchedrin and Zelevinsky}(2011)}]{Sch11}
\bibinfo{author}{\bibfnamefont{G.}~\bibnamefont{Shchedrin}} \bibnamefont{and}
  \bibinfo{author}{\bibfnamefont{V.~V.} \bibnamefont{Zelevinsky}},
  \bibinfo{journal}{preprint arXiv:1112.4919v2}  (\bibinfo{year}{2011}).

\bibitem[{\citenamefont{Grobe et~al.}(1988)\citenamefont{Grobe, Haake, and
  Sommers}}]{Gro88}
\bibinfo{author}{\bibfnamefont{R.}~\bibnamefont{Grobe}},
  \bibinfo{author}{\bibfnamefont{F.}~\bibnamefont{Haake}}, \bibnamefont{and}
  \bibinfo{author}{\bibfnamefont{H.-J.} \bibnamefont{Sommers}},
  \bibinfo{journal}{Phys. Rev. Lett.} \textbf{\bibinfo{volume}{61}},
  \bibinfo{pages}{1899} (\bibinfo{year}{1988}).

\bibitem[{\citenamefont{Fyodorov et~al.}(1997)\citenamefont{Fyodorov,
  Khoruzhenko, and Sommers}}]{Fyo97a}
\bibinfo{author}{\bibfnamefont{Y.~V.} \bibnamefont{Fyodorov}},
  \bibinfo{author}{\bibfnamefont{B.}~\bibnamefont{Khoruzhenko}},
  \bibnamefont{and} \bibinfo{author}{\bibfnamefont{H.-J.}
  \bibnamefont{Sommers}}, \bibinfo{journal}{Phys. Rev. Lett.}
  \textbf{\bibinfo{volume}{79}}, \bibinfo{pages}{557} (\bibinfo{year}{1997}).

\bibitem[{\citenamefont{Sokolov and Zelevinsky}(1989)}]{Sok89}
\bibinfo{author}{\bibfnamefont{V.~V.} \bibnamefont{Sokolov}} \bibnamefont{and}
  \bibinfo{author}{\bibfnamefont{V.~G.} \bibnamefont{Zelevinsky}},
  \bibinfo{journal}{Nucl. Phys. A} \textbf{\bibinfo{volume}{504}},
  \bibinfo{pages}{562} (\bibinfo{year}{1989}).

\bibitem[{\citenamefont{Poli et~al.}(2009{\natexlab{b}})\citenamefont{Poli,
  Dietz, Legrand, Mortessagne, and Richter}}]{Pol09a}
\bibinfo{author}{\bibfnamefont{C.}~\bibnamefont{Poli}},
  \bibinfo{author}{\bibfnamefont{B.}~\bibnamefont{Dietz}},
  \bibinfo{author}{\bibfnamefont{O.}~\bibnamefont{Legrand}},
  \bibinfo{author}{\bibfnamefont{F.}~\bibnamefont{Mortessagne}},
  \bibnamefont{and} \bibinfo{author}{\bibfnamefont{A.}~\bibnamefont{Richter}},
  \bibinfo{journal}{Phys. Rev. E} \textbf{\bibinfo{volume}{80}},
  \bibinfo{pages}{035204(R)} (\bibinfo{year}{2009}{\natexlab{b}}).

\end{thebibliography}

\end{document}